# Ion Sources for Medical Applications

*S. Gammino*
Istituto Nazionale di Fisica Nucleare, Laboratori Nazionali del Sud, Catania, Italy

**Abstract**
Ion sources are key components of accelerators devoted to different types of medical applications: hadron-therapy facilities (accelerating protons or carbon ions), high-intensity accelerators for boron-neutron capture therapy (using intense proton beams), and facilities for isotope production (using different ion species). The three types of application present different requirements in terms of ion beam quality, reproducibility, and beam availability. Different characteristics of ion sources will be described, along with the reasons why they are particularly interesting or largely used.

## 1   Introduction

For medical facilities, ion-source design must be oriented according to the needs of three demanding applications:

–   hadron-therapy facilities (accelerating protons or carbon ions);

–   BNCT (boron-neutron capture therapy, accelerating protons);

–   isotope production (producing different ion species).

The requirements for an ion source devoted to medical applications are not unequivocally determined. High currents are required (between hundreds of μA and tens of mA), but there are more stringent constraints than the one on beam intensity. The characteristics for ion sources devoted to these purposes can be summarized as follows:

a)  ability to produce the necessary beam current for the treatment (with a contingency margin of 20% or more), i.e. 200–400 eμA for carbon ions and three-times more for protons (up to tens of mA of protons in the case of BNCT facilities);

b)  beam emittance lower than accelerator acceptance, typically 0.5-0.7 $\pi$ mm mrad normalized;

c)  high stability, low beam ripple, and high reproducibility;

d)  user friendly;

e)  high MTBF (mean time between failure);

f)  low maintenance.

Requirements d)–f) are very important for all types of application, since hospital facilities need versatile devices with a short tuning time.

Some sources may be adapted for different applications, i.e. a source producing multiply charged ions may work for hadron therapy and for isotope production, while intense beams of protons are used for BNCT and for isotope production. Often, in order to minimize the spare parts and the overall complexity, the same type of ion source is made available in two copies, in order to be interchangeable. Uptime is essential for medical applications, more than for ion sources devoted to nuclear-physics accelerators. The above-mentioned characteristics restrict the scope of possible ion sources to those high-current ion sources based on the formation of a dense and hot plasma. The ions are generated inside the plasma, and then extracted by beams of electrostatic optical elements. The production of the plasma, the extraction of ions from the plasma, and the beam management coincide to fulfil the requirements, and particular care is given to each stage to avoid the emittance growth (and poor transport of the beam to the users) and the beam halo formation.

## 1.1 General technical characteristics

The definition of ion-source properties is given by the mechanism of plasma formation. In a few words, a good knowledge of the plasma parameters simplifies the fulfilment of the requirements. Laboratory plasmas—including the ones generated in compact ion sources—can be produced in several manners, and they differ from each other in the electron temperature, electron density, and plasma lifetime. Electrical discharges under vacuum, electron beams passing through neutral gases or, in a more complicated manner, electromagnetic waves interacting with gases or vapours, in the presence of a well-shaped magnetic field, are the mechanisms used for inducing discharges in gaseous systems, and producing plasmas with suitable characteristics as sources of singly charged or multi-charged ions. Some common features of any ion source used in research or medical laboratories are summarized here:

- a cylindrically shaped under-vacuum cavity with metallic walls, with vapour or gas fluxed before turning on the plasma;
- some flanges for gas feeding, ovens, electron beams, and other tools for plasma discharge tuning;
- a system feeding the energy needed to the discharge (electrostatic energy, electromagnetic waves, etc.);
- a system for plasma confinement (e.g. magnetic trap);
- an ion extraction system able to optically manipulate the beam, ensuring good collimation, low energy spread, and low emittances.

Other ancillary systems include pumps, control systems, and several kinds of diagnostics (for both the plasma and the produced beam). More information is available in [1–3].

Ion sources may be chosen from among a large variety, e.g.:

1) PIG: i.e. Penning ion sources [4];
2) EBISs: i.e. electron beam ion sources;
3) LISs: i.e. laser ion sources;
4) ECRISs: i.e. electron cyclotron ion sources.
5) H-source 'family':
    a) helicon sources;
    b) surface plasma sources;
    c) volume sources;
    d) RF sources.

Sources 1)–3) in the above list are not convenient for the use in medical facilities for several reasons. PIG sources are unfit for medical applications due to their short lifetime. EBISs are more appropriate for nuclear-physics facilities requiring low currents but extremely high charge states. LIS produced beams are affected by high-energy spreads. Sources 4) and 5) (including all the different kinds of sources belonging to this 'family') are therefore the most used in worldwide medical facilities. The motivation of their use and the specifications that are particularly relevant will be hereinafter described, while for the operation principle of each type of ion source, we refer to [1–3].

## 2 Hadron therapy facilities

Because of the increasing interest in different ion species and beam features, and because of the larger social impact of hadron therapy, a large part of this note is devoted to such ion sources.

Hadron therapy, up to now, has used very few types of particles, thus limiting the requirements for ion sources to proton (even for neutron generation) and carbon beams, in most cases. Rarely, deuteron, helium, oxygen, neon and argon beams have been required.

### 2.1 Ion sources for proton therapy

The current rate requested for proton therapy is always in the range of 1 mA or even less, so the types of proton sources [1–3] used as injectors for proton-therapy accelerators are:

– multicusp volume or surface plasma sources for $H^-$;
– 2.45 GHz microwave discharge ion source;
– duoplasmatron;
– PIG ion sources.

Several types of $H^-$ sources are used, based on different mechanisms of plasma generation and beam formation. Surface plasma sources, with magnetrons, Penning ion gauge (PIG) ion sources with and without convertors, as well as magnetic-multipole volume sources, with and without caesium, are used. The methods of igniting and maintaining magnetically confined plasmas may be quite different: hot and cold cathodes, radio frequency and microwave power, etc. The extraction systems are specialized and use magnetic and electric fields to guide the beam towards the low-energy beam transport line, providing cw (continuous wave) beams or pulsed beams with ad hoc time structures.

Simple commercial multicusp sources are often used to produce 1–2 mA of protons [5] while more complex devices are developed for larger currents. Several improvements have been implemented to the $H^-$ sources in the last decades; e.g. at Lawrence Berkeley National Laboratories (LBNL, USA) many efforts were made in developing innovative RF antennas for the optimization of RF matching to the plasma. In 1990, Leung *et al.* [6] reported the use of inductively generated plasma for producing $H^-$ beams 'with almost no lifetime limitation'. The efficiency was shown to be higher than the case of a source with a filament. In 1993, the same group reported [7] a three-fold gain in $H^-$ beam using a collar with a SAES company Cs dispenser. In 1996, Saadatmand *et al.* [8] increased the current to 70–100 mA running at 10 Hz 0.1 ms with the SSC (Superconducting Super Collider) source modelled after the LBNL source. The $H^-$ beam appeared to be stable for up to 8 hrs.

Additional advancements were achieved at LBNL with the SNS (Spallation Neutron Source) $H^-$ ion source: a caesium-enhanced, multicusp ion source [9]. For the ion extraction and transport, an advanced Low-Energy Beam Transport (LEBT) line was designed in order to properly manipulate the extracted beam. The very compact LEBT, shown in Fig. 1, was made of a two-lenses, electro-static system, only 12 cm long. Lens 2 was split into four quadrants to steer, chop, and blank the beam.

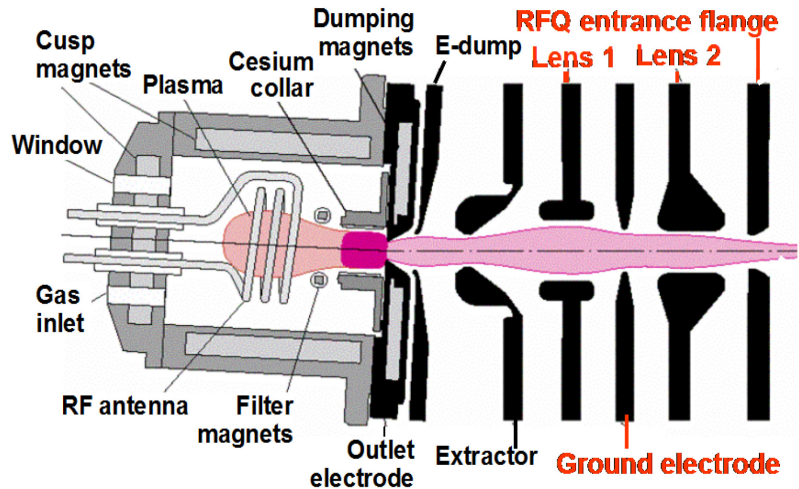

**Fig. 1**: Sketch of the H⁻ source with a compact LEBT [10]

The results are good and the source is suitable with some limitations for medical accelerators: a 30 mA peak current is produced with horizontal r.m.s. normalized emittance of 0.115 mm mrad, and, even when running Cs free, the H⁻ volume source produces 10–15 mA of H⁻.

Another widely used H⁻ source is the so-called Helicon Discharge Surface Plasma Source [10]. The versatility of this source, along with its ability to produce different ion species, the fairly long MTBF, and relatively good stability, might be applicable to different medical facilities. Figure 2 shows the main subsystems of these devices, including: 1–gas valve; 2–discharge volume; 3–discharge vessel; 4–helicon saddle-like antenna; 5–magnetic coil; 6–ion/atom converter; 7–electron flux; 8–emission aperture (slit); 9–extraction electrode; 10–suppression /steering electrode; and 11–ground.

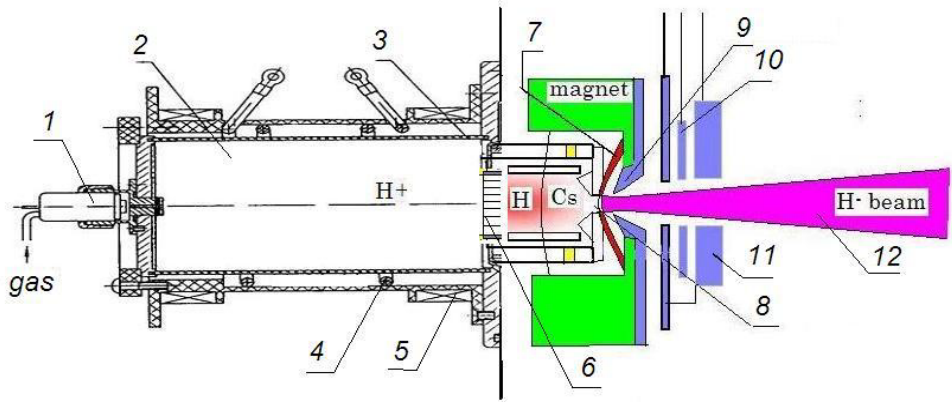

**Fig. 2:** Scheme of helicon discharge surface plasma source

The H⁺ sources which may be suitable for medical facilities are the 2.45GHz off-resonance discharge microwave ion sources (also known as microwave discharge ion sources—MDIS) [11–12]. They present many advantages in terms of compactness, high reliability, ability to operate in cw mode or in pulsed mode, reproducibility, and low maintenance. High-current proton beams may be delivered, up to 100 mA**,** with low transversal r.m.s. normalized emittance, of the order of 0.20 to 0.30 π mm mrad. The major advantage of MDIS comes from the absence of antennas, which makes this equipment reliable for long duration operations (even months) with good reproducibility. This is a condition for computer-controlled operations, without intervention of operators. The scheme of the source is simple and is shown in Fig. 3.

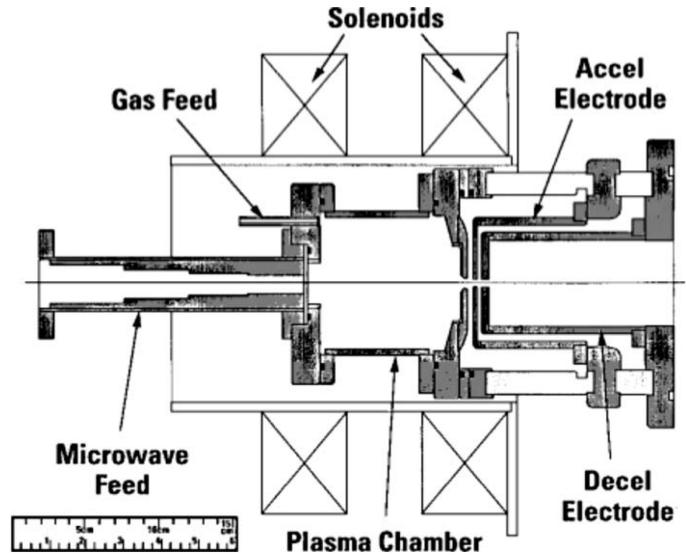

**Fig. 3:** Original MDIS design at Chalk River National Laboratory [11]

## 2.2 Ion sources for carbon therapy

ECRISs, EBISs, and laser ion sources may be able to produce multiply charged ions, but only ECRISs are suitable for carbon therapy (and proton therapy as well), as they can provide beam currents of hundreds of microamperes with good long-term stability and reproducibility.

The operating principle and additional information may be found in [1, 13] along with the scaling laws that determine the beam characteristics. Some conditions have been explored in the 1990s that permit a long enough plasma confinement time to be obtained and multiply charged ion beams to be produced. For the sake of brevity, let us say that the conditions needed to produce intense beams of $C^{4+}$ and $C^{6+}$ used in hadron therapy may be summarized by the high-$B$ mode concept, which states that for a high-frequency Electron Cyclotron Resonance (ECR) source, the magnetic confinement should obey the rule

$$B/B_{ECR} > 2.$$

According to this concept, a higher $B/B_{ECR}$ increases the electron temperature and the ion lifetime and fully stripped heavy ions are obtained. It does not explain the evolution of the electron temperature $T_e$ and its dependence on the power density in the plasma chamber (microwave coupling to plasma is limited by other instabilities in addition to the magnetohydrodynamical ones, which instead limit the ion lifetime). Moreover, the charge exchange process becomes crucial for inner-shell electrons and, even in presence of high production rates of highly charged ions (HCI) in the plasma, few HCI appear in the extracted beam if the pressure is not further improved, in the range of $10^{-7}$ mbar or better. The operating frequency also plays a role and the ECRISs for hadron therapy use microwave generators with frequencies equal to, or higher than, 10 GHz.

The description of the ECRIS operating principles in [1, 13] does not complete the scope of specific requirements of ECRISs for hadron-therapy facilities. In fact, though the conditions of magnetic confinement, vacuum, and microwave frequency are satisfied, the result may be not sufficient for the full exploitation of an accelerator facility for treatment, and additional conditions must be satisfied in order to ensure patient safety and successful treatment. Additional interlocks on the RF system, or high voltage power supplies, permit the medical operator to ensure the patient's safety, while for successful treatment, it is necessary to have a stable emittance figure and a beam ripple as low as possible.

It should be also considered that ECRISs must be adapted to make them suitable for hospital facilities:

- a hospital facility is not adapted to the 'difficult case' of multi-parameter ECR sources that can be managed in a laboratory environment, so the most advanced third generation ECRIS must be discarded [14];
- a 'high-performance conventional ECRIS' has large operating costs for electricity, which is not the case for a source with permanent magnets, even if its performance is worse;
- minimizing the number of components subject to failure makes the source reliability better.

Standard ECR ion sources have been specialized for carbon therapy. The two series most commonly known worldwide are the Supernanogan type and the so-called KEI series. Their main features are listed in Table 1. Figure 4 shows a picture of the two ion sources [15–16].

**Table 1:** Main features of commercial ion sources

| Source | Supernanogan | KEI series |
|---|---|---|
| Type | ECR | ECR |
| Magnets | permanent | Permanent |
| Ion | Carbon | Carbon |
| Charge/current | 4+/200–250 µA | 4+/240–430 µA |
| Extraction voltage | 24 kV | 30 kV |
| Frequency | 14.25–14.75 GHz | 9.75–10.25 GHz |
| Operation | CW | Pulse |
| Gas | $CO_2$ | $CH_4$ |

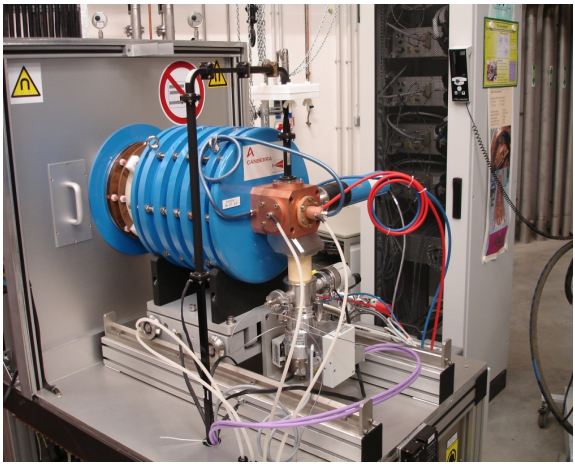 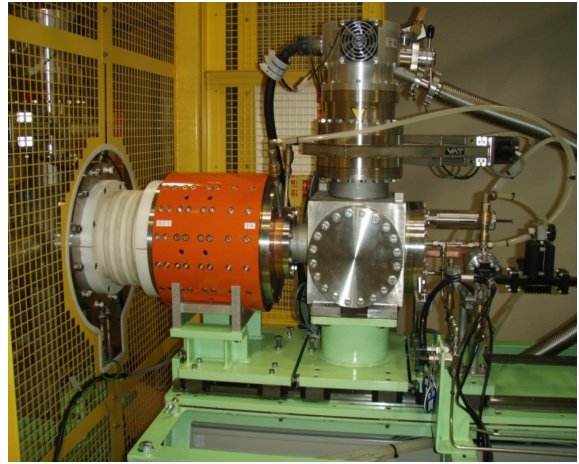

**Fig. 4:** (left) A Supernanogan-type ion source [15], and (right) a KEI-type source [16]

The KEI sources are used in Japanese facilities, whilst the Supernanogan type are largely used in Europe, and they became a standard solution as beam injectors of hadron-therapy facilities, in spite of their limited beam brightness. The Supernanogan source is equipped with a double-wall, water-cooled plasma chamber, with a 7 mm diameter aperture for beam extraction. The permanent-magnet system provides the axial and radial confinement (axial field from 0.4 to 1.2 T, radial field 1.1 T), and the extraction system and RF injection have been optimized through the years for the maximum reliability.

The RF injection consists in the so-called, copper-made 'magic cube', made of a waveguide to coaxial converter with a tuner to minimize the reflected power. An RF window is used for the junction between the magic cube at high vacuum and the waveguide at atmospheric pressure.

The injection flange hosts also a DC bias system to add electrons to the plasma and decrease the plasma potential. An RF generator, of about 400 W at 14.5 GHz, is used for feeding the plasma (the effective power used in operation is below 300 W). In order to make a further optimization of the beam properties (both current and emittance) possible, flexible frequency variable travelling-wave-tube amplifiers (TWTAs) are used in order to exploit the so-called frequency tuning effect (FTE) [17]. The main beam parameters, and the improvements proposed by INFN-LNS scientists, are listed in Table 2 (implementation of a frequency tuning device for the microwave injection; design of a new extraction system for further improvement of the beam emittance and beam stability; and changes to the gas input system in order to achieve a much better stability).

The FTE was applied since significant improvements were observed at INFN-LNS when a source was fed by a klystron or a Travelling-Wave-Tube (TWT) based generator. It was then understood that FTE does not only change the plasma density, but it also strongly affects the beam formation dynamics. In Fig. 5, it is shown that a wide fluctuation of the output beam current is obtained by varying the feeding frequency for a $C^{4+}$ ion extracted by the Supernanogan source of CNAO, Pavia.

The impact of the wave frequency on the formation and spatial distribution of the warm electrons, including the effects on ion dynamics, was investigated in [18] showing that: i) the energy absorption is influenced by the electromagnetic field modal distribution inside the cavity, affecting the heating rapidity; ii) the frequency also impacts on the density distribution—warm electrons are mostly formed where a high field intensity exists; and iii) the RF heating near resonance induces the formation of a non-homogeneously distributed plasma. All this information can be exploited for our purpose, because the variation of the electron energy distribution fraction may optimize the production of some charge states while the electromagnetic mode distribution may improve the beam emittance.

**Table 2:** Main beam parameters and improvement steps of the Supernanogan ion source (made with the aid of INFN-LNS ion-sources team).

| Ions | Current (requested) [µA] | Current (available) [µA] | After improvements by INFN-LNS [µA] | Emittance (requested) [π mm mrad] | Emittance (measured) [π mm mrad] | Stability [99.8%] |
|---|---|---|---|---|---|---|
| $C^{4+}$ | 200 | 200 | 250 | 0.75 | 0.56 | 36 h |
| $H_2^+$ | 1000 | 1000 |  | 0.75 | 0.42 | 2 h |
| $H_3^+$ | 700 | 600 | 1000 | 0.75 | 0.67 | 8 h |
| $He^+$ | 500 | 500 |  | 0.75 | 0.60 | 2 h |

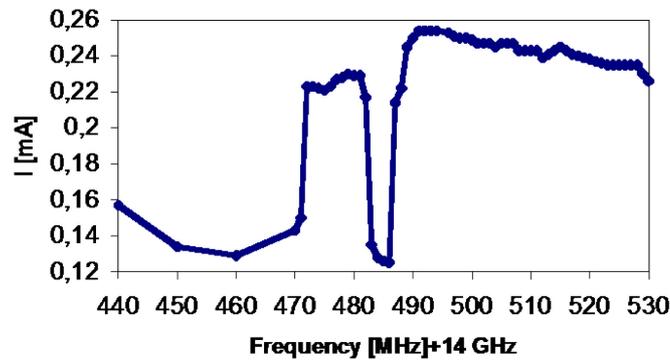

**Fig. 5:** Experimental evidence of frequency tuning effect with the Supernanogan-type ECRIS of CNAO, around 14 GHz [17].

Additional 'alternative heating schemes'—like two-frequency heating (TFH)—have proved very helpful in improving the beam stability, which indeed represents a key challenge for cancer treatment machines. TFH consists of simultaneously launching two electromagnetic waves into the plasma, at different frequencies, in order to improve the plasma confinement and the overall stability. This, in turn, decreases the extracted beam ripple, as can be seen in Fig. 6.

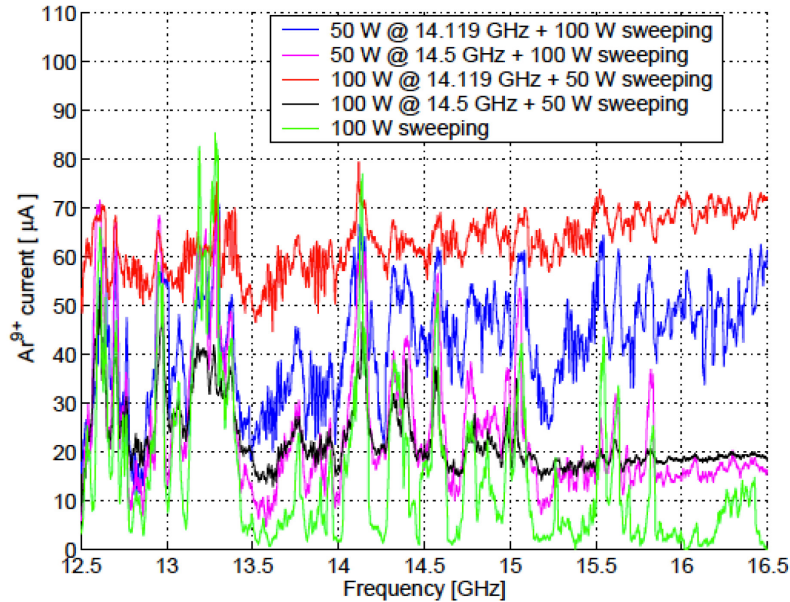

**Fig. 6:** Impact of TFH—two frequency heating in terms of current and beam ripple damping [19]

Future perspectives in hadron therapy—like the production of higher carbon-ion currents and metallic ions as lithium or beryllium—will require more performing ECRISs.

The AISHa (Advanced Ion Source for Hadrontherapy) project [20] will try to overcome such challenges. AISHa is a hybrid ECRIS: the radial confining field is obtained by means of a permanent magnet hexapole, while the axial field is obtained with a set of four superconducting coils. The superconducting system will be helium free at 4.2 K, by using two cryocoolers. The magnetic field values are following the scaling laws (R. Geller) and the high-$B$-mode concept [13–14]. The operating frequency of 18 GHz has been chosen to maximize the plasma density, taking into account the availability of commercial microwave tubes and the specificity of the installation in hospital environments. The electric insulation is chosen to be 40 kV, for daily operation above 30 kV. A sketch of the AISHa ion source is shown in Fig. 7, while the main features are listed in Table 3.

The set of four superconducting coils independently energized will permit a flexible magnetic trap to be realized and the electron energy distribution function (EEDF) to be adapted to the users' needs. The use of a broadband microwave generator able to provide signal with complex spectrum content, will permit the frequency to be tuned efficiently, increasing the electron density, and therefore, the performance, in terms of current and average charge state produced.

The chamber dimensions and the injection system have been designed to optimize the microwave coupling to the plasma chamber, taking into account the need for space to house the oven for metallic ion beam production. This feature may represent a significant step above the current technology, with the goal of keeping a very low beam ripple, even in the presence of larger plasma density and high oven temperature.

The ability of the AISHa source to operate with larger RF power may be applied to the production of $C^{6+}$ for compact cyclotrons. In fact, the use of an accelerator chain based on a synchrotron increases the cost of a heavy-ion-therapy facility, but, up to now, the success of compact cyclotrons accelerating $C^{6+}$ has been limited by the insufficient stability, reliability, and reproducibility of high-performance ECRISs. For example, the HEC (high-voltage extraction configuration) source developed at NIRS-Chiba (National Institute of Radiological Sciences) is a powerful ECRIS operating at 18 GHz, and its performances are more than double the KEI sources, but conversely, its application in a hospital has not been judged as viable by the research group that developed it, because of its complexity and reproducibility.

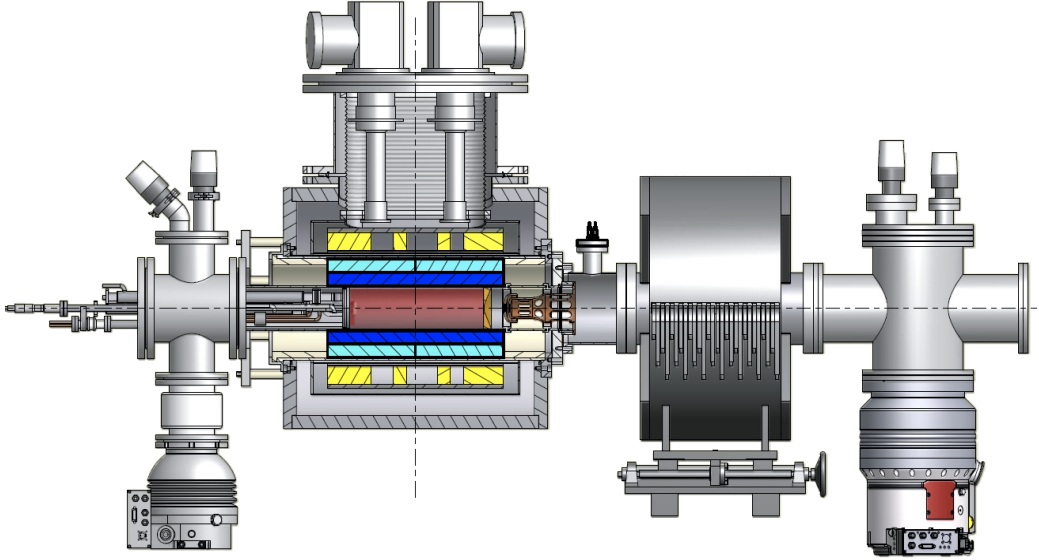

**Fig. 7:** Sketch of the AISHa ion source [20]

**Table 3:** Main operating parameters of the AISHa source

| Parameter | Value |
| --- | --- |
| Radial field | 1.3 T |
| Axial field | 2.6 T–0.4 T–1.5 T |
| Operating frequencies | 18 GHz (TFH) |
| Operating power | 1 kW |
| Extraction voltage | 40 kV |
| Chamber diameter / length | $\Phi$ 92 mm / 300 mm |
| LHe | Free |
| Iron yoke diameter / length | 42 cm / 60 cm |
| Source weight estimation | 480 kg |

## 2.3 Ion sources for BNCT

Ideal ion sources for BNCT should fulfil the requirements of neutron-flux production via reactions induced by intense proton beams. The beam current (with a margin of 20% or more) should lie in the range of many mA of protons. Even in this case, high stability and high reproducibility are mandatory, along with a very high MTBF.

Two types of ion sources are particularly suitable for BNCT facilities: multicusp sources of $H^-$ beams and MDIS for $H^+$.

A typical example is the multicusp source of Kyoto University. The source is able to provide 15 mA of H⁻ beams with 0.66 π mm mrad 4 r.m.s. normalized emittance [21]. The source was developed in 1990 for injection in a TR70 cyclotron. The first version was able to produce up to 7 mA H⁻ beams, but the upgraded version (1994) allowed the source performances to improve by up to 15 mA cw at 5 kW of arc power.

MDIS ion sources can be useful sources for BNCT since they generate beams characterized by low emittance (< 0.2 π mm mrad) and beam ripple. They are able to guarantee very high MTBF due to their reliability. Figure 8 shows the Trasco Intense Proton Source (TRIPS), a high-intensity microwave source, whose goal is the injection of a maximum proton current of 35 mA in a radio-frequency quadrupole (RFQ), with an r.m.s.-normalized emittance lower than 0.2 π mm mrad at the operating voltage of 80 kV [22]. TRIPS was originally developed as a proton source for driving a subcritical reactor to transmute nuclear waste, but it is considered for BNCT facilities at INFN. It has been tested for reliability, which reached an interesting value of 99.8% over 142 h.

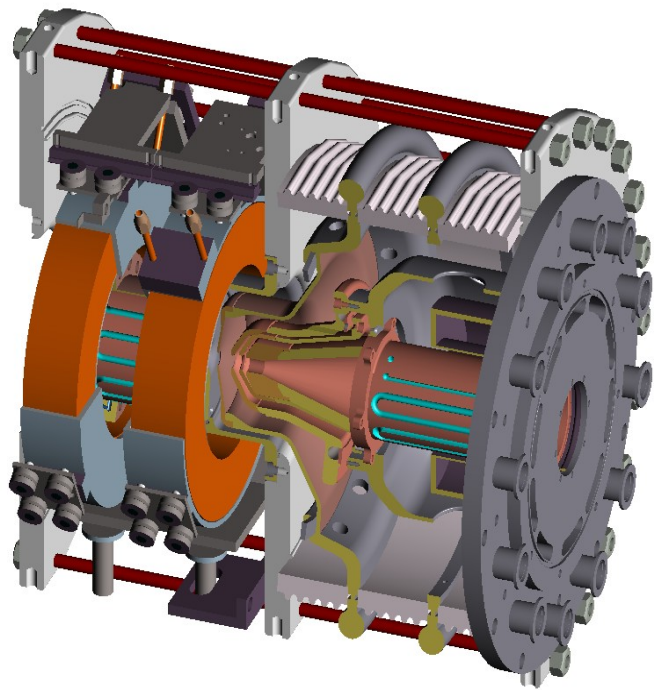

**Fig. 8:** Layout of the TRIPS proton source designed and tested at INFN-LNS [22]

## 2.4 Ion sources for isotope production

Ion sources for isotope production must be characterized by beam currents as large as possible (depending on the limits on target reliability) and emittance lower than the accelerator acceptance. Also, in this case, a high MTBF is required. Commercial cyclotrons for the in-situ production of protons or deuterons at energies in the range 10–100 MeV, with beam currents up to 2 mA, are available.

H⁻ ion sources are often used, such as a multicusp ion source, or RF-driven ion sources and helicon sources. Other options, such as ECR-driven ion sources, have been considered but they have a narrow application because of the larger cost and complexity.

It can be seen that the development of sources for isotope production is not subject to breakthrough but is rather the outcome of technological steps. For example, the development of the filament-driven surface conversion ion source regarded the filament properties. In fact, the filament strength and longevity depend on the material grain size, impurities, and processing. For example, adding 3% of rhenium into tungsten filaments has been proven to enhance the material properties in other applications.

The improvement of the filament material would allow longer lifetimes at the same performance level or higher currents at the same lifetime. The improvement of the source temperature control permits higher performance, provided that the lifetime does not change significantly. Furthermore, the improvements in beam current allow the size of the extraction aperture to be reduced, leading to the decrease of the emittance.

The feasibility of RF-driven $H^-$ ion sources has been attractive because of their longer lifetimes, obtained by means of an external antenna, and their larger plasma density. Furthermore, surface production is more effective and technologically simpler [23]. The neutral pressure seems to be the limiting parameter in the case of the helicon source, whilst the 2.45 GHz MDIS could be a useful kind of source for overcoming any limitations in future years.

## 3   Conclusions

In this presentation, I have outlined some peculiar aspects of the science and technology of ion sources, rather than the working principles of the different types of ion sources which cannot be described in a short paper; they are available in the CERN Accelerator School textbook and in other relevant textbooks [1–3]. More relevance was given to the comprehension of the requirements set by the different medical applications along with the ability to provide the needed specifications of the source over a long time period, with decent reproducibility in order to guarantee the satisfaction of the users, particularly important in the case of therapy. There is plenty of room for future improvements coming from a better knowledge of the basic processes and improved ability to simulate them and to forecast the produced beam properties. The development of better ion sources may permit the success of treatment to be increased, the cost to be decreased, and finally the application of accelerators to be extended to medical facilities as the result of increased social awareness. The optimization of performance of ion sources and their matching to the accelerators may improve the budget of the isotope production facilities and may extend the variety of isotopes with great advantages for medical diagnostics.

## Acknowledgement

I wish to thank my colleague David Mascali for the editing and final revision of this paper.